\documentclass[12pt,preprint]{aastex}
\input epsf.tex

\begin{document}
 
\title{THE LOWER MAIN SEQUENCE AND MASS FUNCTION OF THE GLOBULAR CLUSTER MESSIER 4$^1$}

\author{
Harvey B. Richer\altaffilmark{2},
James Brewer\altaffilmark{2},
Gregory G. Fahlman\altaffilmark{3},
Brad K. Gibson\altaffilmark{4},
Brad M. Hansen\altaffilmark{5,6},
Rodrigo Ibata\altaffilmark{7},
Jasonjot S. Kalirai\altaffilmark{2},
Marco Limongi\altaffilmark{8},
R. Michael Rich\altaffilmark{5},
Ivo Saviane\altaffilmark{9},
Michael M. Shara\altaffilmark{10} \&
Peter B. Stetson\altaffilmark{11}   }

\shorttitle{M4 Main Sequence}
\shortauthors{Richer {\it et al.}}

\altaffiltext{1}
{Based on observations with the NASA/ESA Hubble Space Telescope, obtained at
the Space Telescope Science Institute, which is operated by AURA under NASA
contract NAS 5-26555. These observations are associated with proposal GO-8679.}
\altaffiltext{2}
{Department of Physics \& Astronomy, University of British Columbia,
6224 Agricultural Road, Vancouver, BC V6T 1Z1, Canada. richer@astro.ubc.ca,
jbrewer@astro.ubc.ca, jkalirai@astro.ubc.ca}
\altaffiltext{3}
{Canada France Hawaii Telescope, 65-1238 Mamalahoa Highway, Kamuela, HI 96743. fahlman@cfht.hawaii.edu}
\altaffiltext{4}
{Centre for Astrophysics and Supercomputing, Swinburne University, Mail 31, 
P.O. Box 218, Hawthorn, Victoria, 3122 Australia. bgibson@astro.swin.edu.au}
\altaffiltext{5}
{Department of Physics \& Astronomy, University of California at Los Angeles,
Math-Sciences 8979, Los Angeles, CA 90095-1562. hansen@astro.ucla.edu, 
rmr@astro.ucla.edu}
\altaffiltext{6}
{Hubble Fellow, Department of Astrophysical Sciences, 122 Peyton Hall, Princeton University, Princeton NJ 08544-1001.}
\altaffiltext{7}
{Observatoire de Strasbourg, 11 rue de l'Universite, F-67000 Strasbourg, France. ibata@newb6.u-strasbg.fr}
\altaffiltext{8}
{Osservatorio Astronomico di Roma, Via Frascati, 33, Rome, I-00040, Italy. marco@nemo.mporzio.astro.it}
\altaffiltext{9}
{European Southern Observatory, Alonso de Cordova 3107, Vitacura, Casilla 19001,
Santiago 19, Chile. isaviane@eso.org}
\altaffiltext{10}
{American Museum of Natural History, Astrophysics Department, Central Park West \& 79th Street, New York, NY 10024-5192. mshara@amnh.org}
\altaffiltext{11}
{National Research Council, Herzberg Institute of Astrophysics, 5071 West
Saanich Road, RR5, Victoria, BC V9E 2E7, Canada. peter.stetson@nrc.ca}

 \vspace{.2in}

\begin{abstract}
The deepest optical image ever in a globular star cluster, a Hubble Space Telescope (HST) 123 orbit exposure in a single field of Messier 4,
was obtained in 2 filters (F606W, F814W) over a 10 week period in early 2001.  
A somewhat shallower image obtained in 1995 allowed us to select out cluster and
field objects via their proper motion displacement resulting in remarkably clean
color-magnitude diagrams that reach to $\rm V = 30$, $\rm I = 28$.
The cluster main sequence luminosity function contains
very few stars fainter than $\rm M_V = 15.0$, $\rm M_I = 11.8$ which, in both filters, is more than 2 magnitudes brighter than our limit. This is
about the faintest luminosity seen among field Population II subdwarfs of
the same metallicity. However, there remains a sprinkling of
potential cluster stars to lower luminosity all the way down to our limiting magnitudes. These latter objects are significantly redder than any
known metal--poor field subdwarf. 
Comparison with the current
generation of theoretical stellar models implies that the masses 
of the lowest luminosity cluster stars observed are 
near $\rm 0.09 M_{\odot}$. We derive the mass function of the cluster in
our field and find that it is very slowly rising towards the lowest masses
with no convincing evidence of a turnover even below $\rm 0.1 M_{\odot}$.
The formal slope between 0.65 and 0.09$\rm M_{\odot}$ is $\rm \alpha = 0.75$ (Salpeter = 2.35) with a 99\% confidence interval 0.55 -- 1.05. A consistency check between these slopes and the number of observed cluster white dwarfs (WDs) yields a range of possible conclusions, one of which is that we have indeed seen
the termination of the WD cooling sequence in M4.

\end{abstract}


\section{Introduction}

Analysis of the Hubble Deep Field (HDF) (Williams {\it et al.} 1995), which is to date
the deepest optical image ever taken, has produced a remarkable series of
important results both in Galactic and extragalactic astronomy (see Ferguson {\it et al.} 2000 for a summary).
With the original HDF as a model, we proposed to carry out an
HDF--like project in the nearest globular cluster, M4. We were successful in obtaining a roughly similar amount of telescope time (123 orbits) for the 2 filters 
F606W (V -- 127,400 sec) and F814W (I -- 192,400 sec) as had been granted for the HDF. Our plan was to reach
$\rm V \sim 30$, $\rm I \sim 28$ with s/n about 3 in each filter in a single M4 field located 6 core radii ($\rm \sim 5\arcmin$) from the cluster 
center -- the same 
field which we observed
in 1995 (GO 5461). 

The main science driver was to search for the oldest and coolest white dwarfs (WDs) in 
M4. The discussion of the WD population in M4 is contained in the
following {\it Letter} (Hansen {\it et al.} 2002).
Other science goals are to attempt to identify the termination of the hydrogen burning MS and hence set the location of the brown dwarf boundary for 
metal--poor
stars, establish the MF in the cluster down to very low masses and
use the cluster WDs to extend the MF beyond the cluster turnoff, investigate the binary frequency in the cluster both through the location
of stars in the CMD and via variability, explore for planets through stellar
occultations, and examine the inner halo populations (Rich {\it et al.} 2002, in preparation) and compare it with
that of the cluster.

\section{The Data and Its Reduction}

The complete data set consists of 15$\times$2600 sec exposures in F555W, 9$\times$800 sec in F814W (1995 -- part of GO 5461 from 
cycle 4 -- see Richer {\it et al.}\ 1995, 1997; Ibata {\it et al.}\ 1999) together with 
98$\times$1300 sec in F606W and 148$\times$1300 sec obtained in F814W (2001 --  
GO 8679 in cycle
9). All images were secured with the same roll angle in the same field of M4 located at about 6 core radii from the cluster center.
We also processed and reduced the small set of F814W images from GO 8153 but in the final reductions
they were not used.

The new images as well as the earlier frames were preprocessed according to the recipes given
in Stetson (1987) and Stetson {\it et al.}\ 1998.   
The individual frames were transformed to the coordinate system of
the first new epoch frame using the MONTAGE2 program. Once the frames
were all on the same coordinate system, those {\it corresponding to a
particular epoch and filter} were averaged together with pixel rejection --
the n highest pixels being rejected to eliminate cosmic ray contamination.
From the HST WFPC2 Manual the mean number
of pixels on a given chip affected by a cosmic ray hit in an 1800 sec exposure
is about 20,000. Hence in a stack of N$\times$1300 sec exposures we expect 0.02257N
pixels at a given position to have suffered a hit. Taking the dispersion
to be $\sqrt{0.02257N}$ and using $5\sigma$ rejection we find that in
combining the 98 F606W images the highest 7 pixels should be rejected while
for the 148 F814W images we should eliminate the highest 9.
Statistically this approach is superior to a median of the frames, which is equivalent to a mean of 2/3 of the number 
of frames, whereas in our procedure
we used 93\% of the images in constructing the mean.

For each chip there are 4 combined frames, $\rm V_{new}$, $\rm V_{old}$, $\rm I_{new}$ and
$\rm I_{old}$. A starlist was
generated for each chip using DAOFIND on the $\rm I_{new}$ and $\rm V_{new}$ frames. We 
then (1) visually inspected each object in the list and rejected those
objects that corresponded to detections of non-astronomical sources
(such as diffraction spikes), (2) subtracted a spatially smoothed
copy of the image from the image itself and inspected this image
for missed stars, and finally (3) combined the 2 lists into our final
starlist.

Stars on each of the chips within the 4 frames were photometered using ALLSTAR
with the starlist as input. ALLSTAR fits a PSF to each object 
in the starlist and does recentering
so as to achieve the best fit. Note that the ability of ALLSTAR to
recenter is essential as this allows shifts between stars on the frames to be
measured which is critical when applying the starlist from the new frames
onto the first epoch images as the proper motion of the cluster relative to the background
amounts to
about 1 HST pixel over the 6 year time baseline. 
To ensure that the individual frames were registered {\it on}
the cluster stars, through an iterative process, we built a transformation
file using only
cluster stars (about 240 per chip) which move insignificantly relative to each other as the observed 
cluster velocity dispersion is 3.5 km/s (Peterson {\it et al.}\ 1996). This 
amounts to a total internal mean motion of about 0.02 pixels over the 6 years.
To compensate for distortions in
the HST optics, we used a 20-term transformation equation. This is in lieu of but
equivalent to applying the Trauger {\it et al.}\ 1995 corrections.
To improve astrometric accuracy, we expanded
the pc pixels by a factor of 2 and the wf pixels by a factor of 3 using
MONTAGE2. This process is similar to ``drizzling'' 
(Fruchter \& Hook 2002). The
individual frames were then averaged together with high pixel
rejection as previously described. It was on these combined, expanded frames
that the final photometry was carried out using ALLSTAR. No charge transfer corrections were made as the high background and the
uniform distribution of the faint stars across the chips suggested that this
was not a serious problem on these images. 

The resulting proper motion displacement (PMD) diagram (Figure 1) separates cluster stars from
the field population rather cleanly.
The clump of non-cluster stars in the PMD diagram clearly has a 
larger dispersion than that of the cluster; with an internal velocity
dispersion of only 0.02 pixels in 6 years, motions within M4 are
unresolved to our level of precision.  
For stars in the
V magnitude range 21 to 26, the measured PMD dispersion for the
cluster is 0.082 pixels ($\rm 0\farcs0082$) which we take as our
positional measurement error (confirmed by simulation) while
the field clump has a dispersion of 0.277 pixls ($\rm 0\farcs0277$).
A complete discussion of this
population will be given in Rich {\it et al.} (2002, in preparation).

In the lower two sections of Figure 1 we display the CMDs for data in different areas of the PMD diagram. The data
were transformed to Johnson V and Kron-Cousins I using stars measured in GO 5461 and calibrated with
ground-based data as discussed in Richer {\it et al.}\ 1997. This resulted in a significant color term in the transformation of F606W to Johnson V. The transformation equations varied slightly from chip-to-chip but for wf2
it was $\rm V = F606W + 0.268(\pm 0.014)(F606W - F814W) + 9.401(\pm 0.031)$.
Photometric measurement errors alone for MS stars as returned by ALLSTAR
ranged from 0.02 magnitudes at $\rm V = 25$ (0.01 at $\rm I = 22$)  to 0.07 at 29 (0.04 at $\rm I = 24.5$).
This resulted in $\rm 1\sigma$ errors in magnitude  and color  for a MS star at $\rm V = 29$, $\rm V-I = 4.5$ of 0.10 and 0.12 respectively.

The left panels
plot all the objects (stars as well as possible galaxies) measured from all 4 chips with no rejection criteria. The central panels isolate the cluster
stars -- objects within 0.5 pixel in radius ($\rm 0\farcs05$) of the mean motion of the cluster.
Our simulations suggest that $\rm <1$\% of cluster stars will have measured PMDs outside this radius. The right panels contain the inner halo stars, again isolated with a 0.5 pixel
radius circle centered on their mean motion with respect to the cluster.
Any faint unresolved galaxies will appear in this CMD.

\section{The  Approach to the Hydrogen-Burning Limit}

M4 is ideally suited for studies of the pop~II hydrogen burning limit, by virtue of its proximity and
relatively (amongst globular clusters) low concentration. From the ground, the best attempt to study the lower
main sequence was made by Kanatas {\it et al}. 1995, who were limited primarily by field star contamination.
Deep images using NICMOS and WFPC on HST allowed Pulone, de Marchi \& Paresce 1999 
to probe estimated masses down to  $\rm \sim 0.15 M_{\odot}$, but with poor statistics. A second epoch of HST observations
allowed Bedin {\it et al}. 2001 to separate cluster members from background stars using the cluster
proper motion, with the faintest cluster members measured at $\rm F814W \sim 24.5$
with masses near $\rm 0.1 M_{\odot}$. Bedin {\it et al}. determined a 
completeness-corrected
luminosity function, finding that their last bin at $\rm I \sim 24.5$ was
statistically indistinguishable from zero. They note that if a 
larger
number of stars are detected and this bin stays empty, then it 
would
strongly indicate the location of the hydrogen burning limit. 
The considerably deeper observation described above allows
us to examine this question in more detail.

Inspection of Figure~1 shows that, while there is indeed a paucity of stars at $\rm V > 27.5$, $\rm I>23.8$, there
does remain a trail of potential MS stars extending redwards, almost to $\rm V = 30$.  These stars are present even with very restrictive proper motion cuts and
have a high probability of cluster membership (Figure 2) so we are convinced that the LF does not have
an abrupt termination bluer than $\rm V - I = 5$. Indeed, there is little reason to expect that it will as the current generation of low mass MS models (Baraffe {\it et al.}\ 1997, Cassisi {\it et al.}\ 2000, Montalban {\it et al.}\ 2000) all predict large changes in 
luminosity
and color with very small changes in mass near the hydrogen burning limit.  Hence
the approach to the hydrogen-burning mass limit should be characterised
by an extended LF, such as is indeed seen in M4.

The least luminous MS stars observed in M4 are significantly fainter than any other known subdwarf at this metallicity (Leggett {\it et al.}\ 1998a; 
compare with LHS 1742a and LHS 377). An immediate consequence is that
there should be correspondingly faint red subdwarfs in the field waiting to be discovered, although they will be rare. For every
field subdwarf in the range $\rm 7.5<M_V<15$, there should be $\sim 0.02$~subdwarfs with $\rm 15<M_V<17.4$, or less
than 1 in 30,000 of all stars in the Solar neighbourhood.

There is also a small sample of apparent cluster members located between the cooling sequence and the MS, and
some are fainter in V than even the least luminous WD detected. It is possible that these are field
objects with proper motions similar to the cluster. If they are cluster members, they could be low luminosity He WDs
in binaries with extremely low mass MS stars. Indeed, such objects are predicted to be the remnants of cataclysmic
variables currently accreting at very low rates (Townsley \& Bildsten 2002). Alternatively, these may indicate
a range of colors for pop~II stars near the hydrogen-burning limit, since low mass, low metallicity stars can be
quite blue (Saumon {\it et al}. 1994).

\section{The Cluster Mass Function}

The derivation of a MF  from the LF for M4 
is an uncertain procedure because the current theoretical models for low temperature stars of this
metallicity do not fit the lower 
MS very well. Modelling difficulties are due to problems in treating convection and opacity in the optically thin, dense and cool outer atmospheres of these stars where molecules dominate and where non-ideal gas effects are important.
This affects both the color-T$_{\it eff}$ relations and the stellar luminosity.
At the lowest masses severe problems with calculating the adiabatic gradient
occur with resultant uncertainty in the mass-luminosity relation (Montalban {\it et al}. 1999).
With these caveats, we generated and display the cluster MF in Figure~3  using the
models of Montalban {\it et al.} to convert luminosities to masses. The input luminosity function was corrected for incompleteness by inserting in 4 trials a total
of 4900 stars with magnitudes from $\rm V=20$ to 31 to each chip. For an artificial star to be considered recovered it had to be found on at
least 3 of the frames (so that its motion could potentially be determined) as well as being recovered within 1 magnitude of its input brightness and within 0.5 pixels of its
input position.

The resulting MF rises very slowly to low masses, with a formal slope
$\rm \alpha = 0.75$ ($\rm N(m) \propto m^{-\alpha}$, Salpeter $\rm \alpha=2.35$). A $\rm \chi^2$ fit finds that slopes of $\rm \alpha = 0.55$ -- 1.05 are acceptable at the 99\% confidence level, 
a tighter constraint will be possible when better
lower MS models are available. There
is no evidence that the MF turns over, even to the lowest masses we detect ($\rm \sim 0.09 M_{\odot}$). This
assertion is more solid than might appear from the uncertain model fitting. The total number of stars
fainter than $\rm I\sim 21$ (brighter than this the models do fit the data) is consistent with the expectations
based on the extrapolated MF down to $\rm M \sim 0.09 M_{\odot}$, regardless of the M-L relation. We can also predict the number
of MS stars expected to $\rm 0.08 M_{\odot}$ (the brown dwarf limit for solar metallicity). Extrapolating the current MF suggests that
we should find 20 more stars redder than $\rm V - I = 5$. If the hydrogen-burning
limit for M4 is at higher mass, {\it e.g.} at $\rm 0.086 M_{\odot}$ as suggested by the Montalban {\it et al}. models,
this will drop to 8. Within this context it is of interest to note that we found no stars with I magnitudes in the range 25 -- 27 that  possess PMDs
appropriate for cluster membership and are found solely on the I frames.

The WD population also allows us to constrain the MF slope above the cluster
turnoff as discussed in Richer {\it et al}. 1997. The completeness-corrected WD counts are 602 to $\rm V=30$. 
The detected 570 MS stars between $\rm 0.09 M_{\odot}$ and $\rm 0.65 M_{\odot}$ (upper limit due to saturation)
imply 1219 WDs if the slope $\rm \alpha=0.75$ extends all the way through the 
turnoff mass ($\rm 0.8 M_{\odot}$) to the expected upper limit for WD progenitor masses ($\rm 8 M_{\odot}$).
Thus, we find a factor of two less than expected. However, if the MF slope is at the steep end
of the acceptable range found above, then we predict 584 WDs, in agreement with the observed numbers. However, the WD counts are far above that expected if the global M4 IMF of
Pulone {\it et al}. 1999 is adopted. They propose $\rm \alpha \sim 2.4$ above the turnoff mass which would
predict a paltry 264 WDs. Furthermore, if M4 has undergone significant mass segregation as
they suggest, this number would be even smaller.  

If there is a significant fraction of the cluster WD population which possess pure He atmospheres,
they will have cooled beyond our detection limit (Hansen 1999) and will not be included in the above census. A population
comprising 50\% of all cluster WDs would allow a continuous power law $\alpha=0.75$ from $\rm 0.09 M_{\odot}$
to $\rm 8 M_{\odot}$. 
Hence the
constraints from WD counts that can be placed on the MS are 
dependent
upon assumptions about the shape of the MF and the WD fraction with He atmospheres. However, the fact that 
all
of the observed WDs seen can be accounted for with a MF slope of
$\rm \alpha \sim 1$ is at least consistent with a simple picture where most 
WDs
are H-rich, and where we have indeed seen the termination of the WD cooling
sequence in M4.

\begin{acknowledgements}
HBR and GGF are supported in part by the Natural Sciences
and Engineering Research Council of Canada. HBR extends his appreciation to the Killam Foundation and
the Canada Council for the award of a Canada Council Killam Fellowship. RMR, MS and IS acknowledge support from proposal GO-8679 and BH from a Hubble Fellowship
HF-01120.01 both of which were provided by NASA through a grant from the Space Telescope Science Institute which is operated by AURA under NASA contract NAS5-26555. BKG acknowledges the support of the Australian Research Council through its Large Research Grant Program A00105171.
\end{acknowledgements}

\newpage

\figcaption[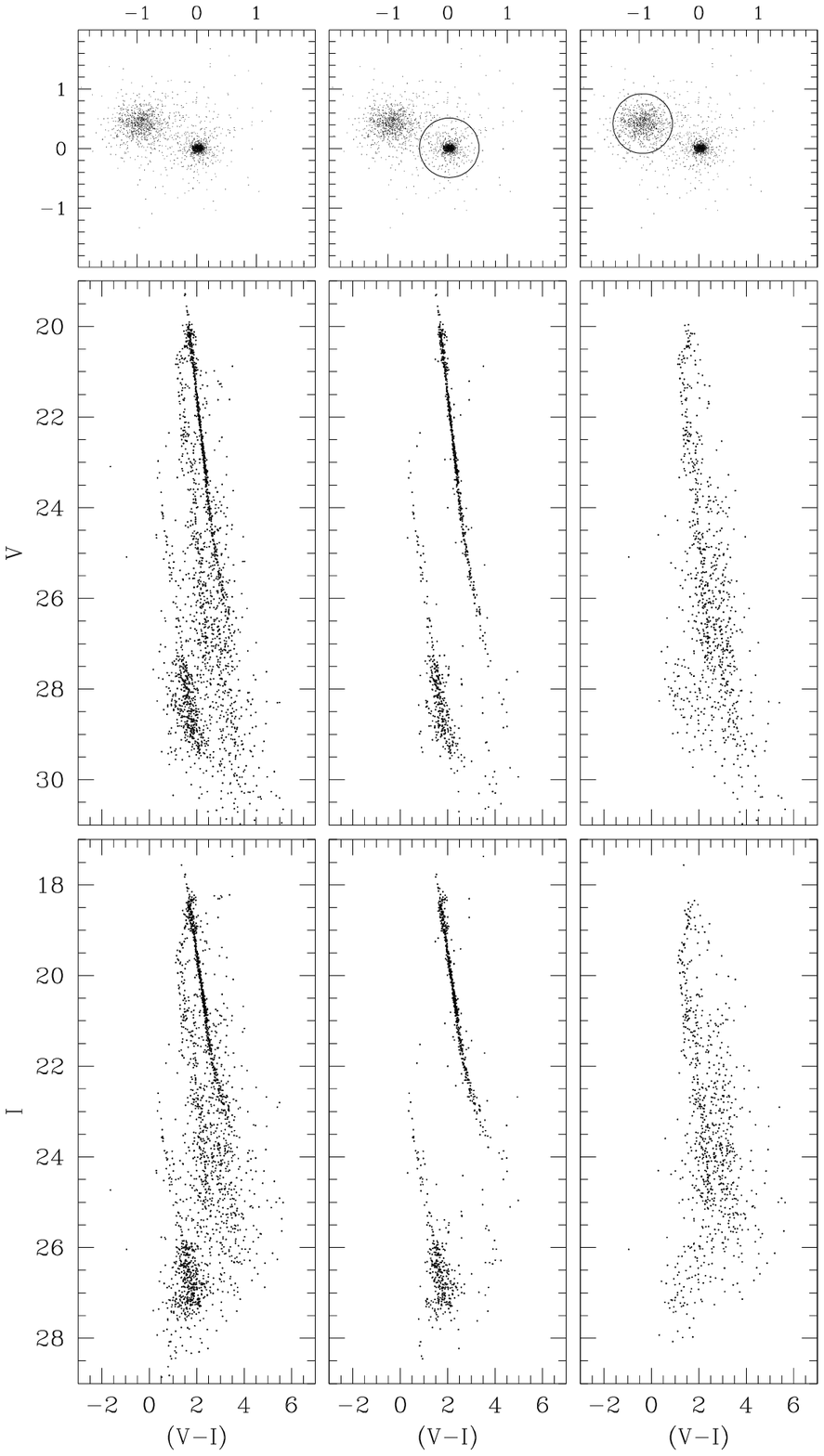]{Upper -- The PMD diagram centered on the cluster.
Each unit is 1 HST pixel. 
Middle -- The V, V$-$I CMDs for objects selected by PMD. The central panel
is for M4 using stars 
with PMDs within
$0\farcs05$/6 yrs of that of the mean motion of the cluster.  Bottom --  
The I, V$-$I CMDs. In the center, the hook to the blue in the WD cooling sequence of M4 is
as predicted by theory and is caused by $\rm H_2$ opacity. \label{CMD}}

\figcaption[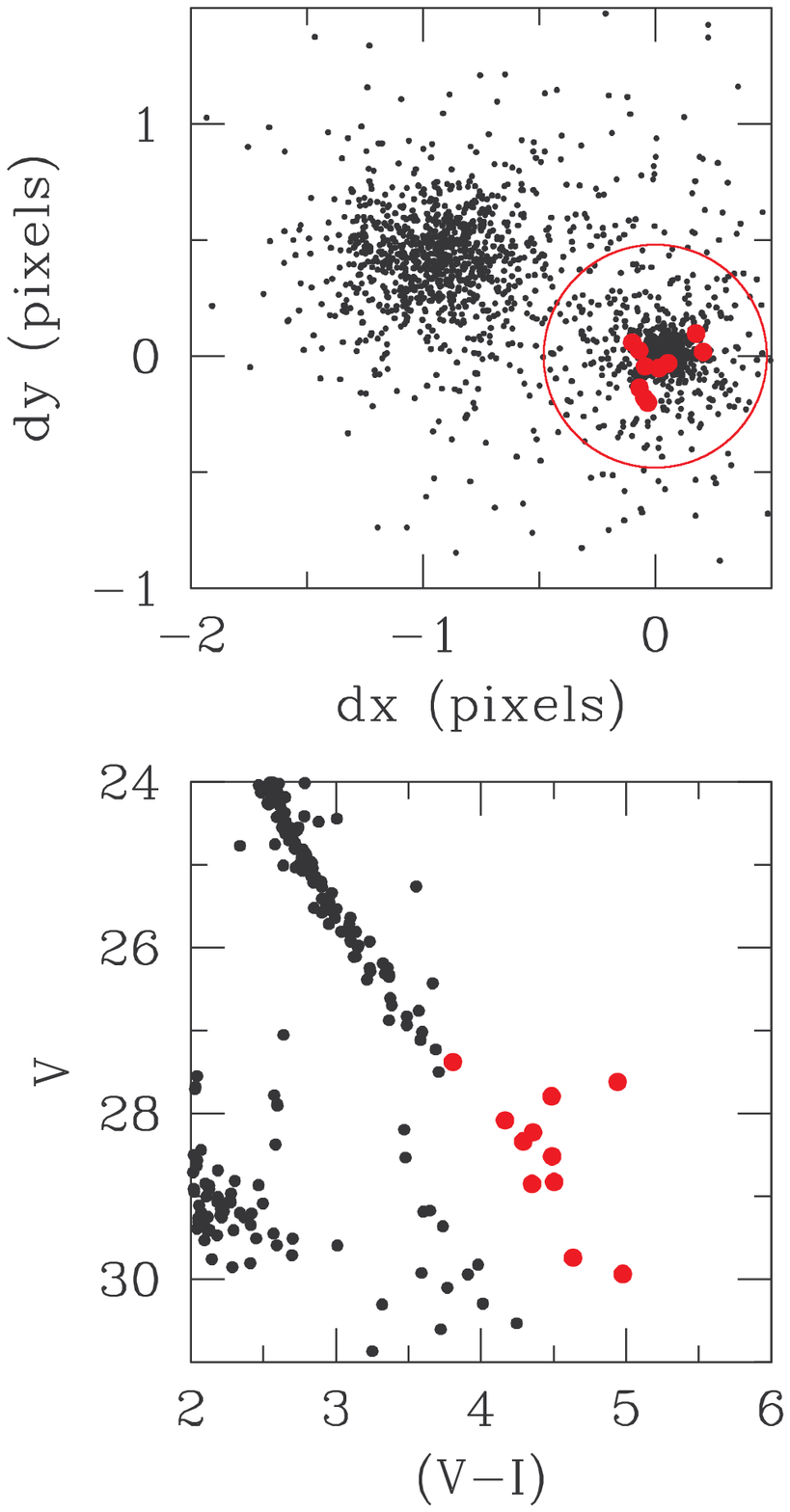]{Upper -- The PMD diagram highlighting with red dots all
those objects
with $\rm V-I > 3.8$, $\rm I < 25.5$ and motion within $0\farcs05$/6 yrs of that of the mean cluster PMD. This illustrates that there is a component of
extremely red stars that, based on their motion,
are very likely cluster members. Bottom -- An expanded version of the PMD selected CMD indicating that these very red objects provide a natural extension to faint magnitudes of the cluster main sequence. All these stars are redder and fainter than any known field subdwarf at the metallicity of M4 (see Leggett {\it et al.} 1998b). \label{PMD}}

\figcaption[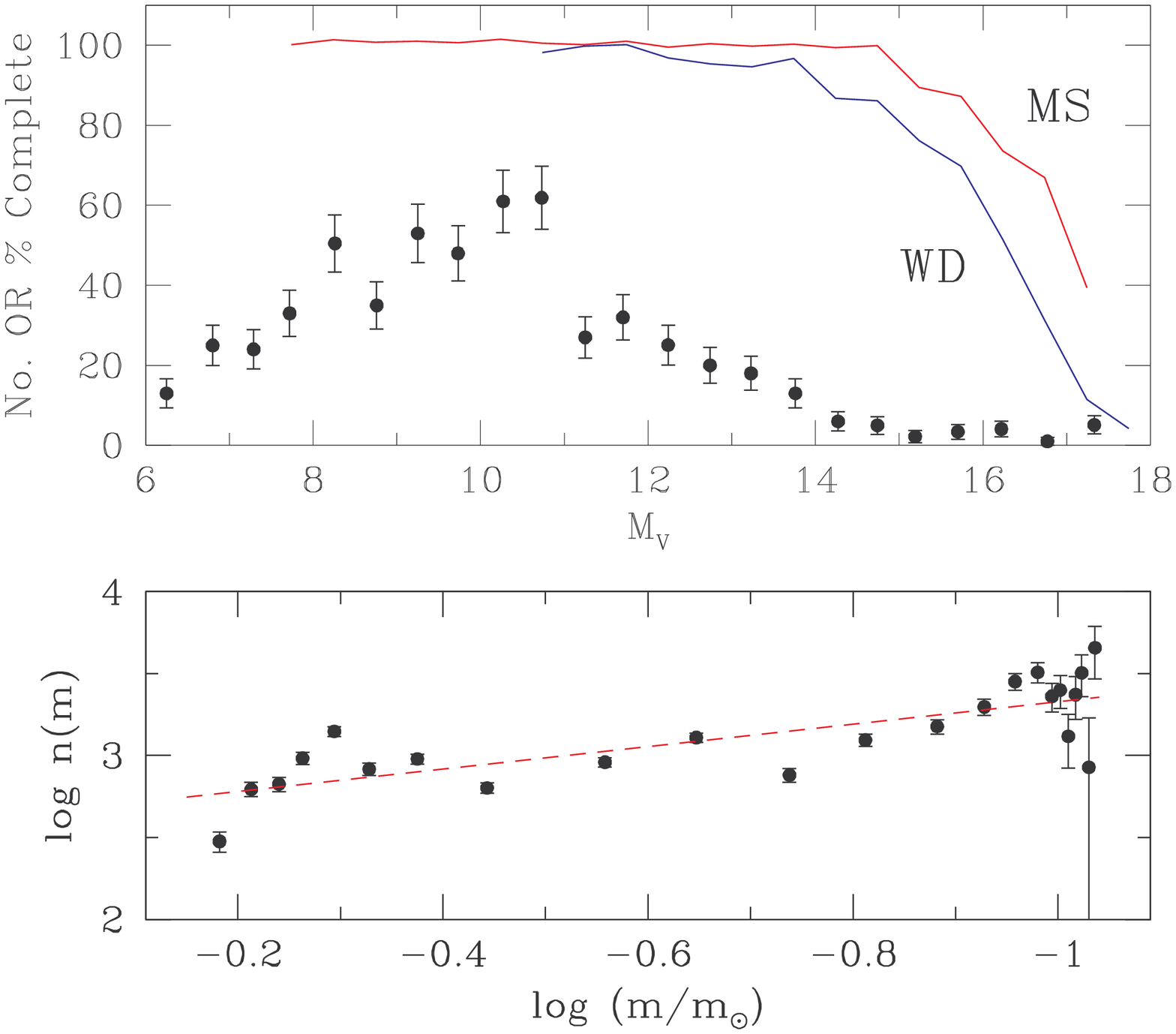]{Upper -- The M4 MS LF corrected for incompleteness and showing these corrections for both MS and WD stars. Lower -- The M4 MS MF with power law slope $\rm \alpha = 0.75$ indicated. \label{MSLF}}

\newpage
\epsscale{0.90}
\plotone{f1.eps}

\newpage
\epsscale{1.0}
\plotone{f2.eps}

\newpage

\plotone{f3.eps}

\end{document}